\documentclass[12pt,preprint]{aastex}
\slugcomment{}
\shorttitle{Gravitationally Lensed Quasar: SDSS J1332+0347}
\shortauthors{Morokuma et al.}
\begin{document}
\title{Discovery of a Gravitationally Lensed 
Quasar from the Sloan Digital Sky Survey: SDSS J133222.62+034739.9
\footnote{Based in part on data collected at Subaru Telescope, 
which is operated by the National Astronomical Observatory of Japan.}}
%
\author{
Tomoki Morokuma\altaffilmark{1},
Naohisa Inada\altaffilmark{1,2}, 
Masamune Oguri\altaffilmark{3,4}, 
Shin-Ichi Ichikawa\altaffilmark{5}, 
Yozo Kawano\altaffilmark{6}, 
Kouichi Tokita\altaffilmark{1}, 
Issha Kayo\altaffilmark{6}, 
Patrick B. Hall\altaffilmark{4,7}, 
Christopher S. Kochanek\altaffilmark{8}, 
Gordon T. Richards\altaffilmark{9,10}, 
Donald G. York\altaffilmark{11}, 
and Donald P. Schneider\altaffilmark{12}
}

\altaffiltext{1}{Institude of Astronomy, School of Science, University of 
Tokyo, 2-21-1, Osawa, Mitaka, Tokyo 181-0015, Japan}
\altaffiltext{2}{JSPS Research Fellow}
\altaffiltext{3}{Kavli Institute for Particle Astrophysics and Cosmology, 
Stanford University, 2575 Sand Hill Road, Menlo Park, CA 94025, USA}
\altaffiltext{4}{Department of Astrophysical Sciences, 
Princeton University, Princeton, NJ 08544, USA}
\altaffiltext{5}{National Astronomical Observatory, 2-21-1 Osawa, Mitaka, 
Tokyo 181-8588, Japan}
\altaffiltext{6}{Department of Physics and Astrophysics, 
Nagoya University, Chikusa-ku, Nagoya 464-8602, Japan}
\altaffiltext{7}{Department of Physics and Astronomy, York University, 4700 Keele Street, 
Toronto, Ontario, M3J 1P3, Canada}
\altaffiltext{8}{Department of Astronomy, The Ohio State University, 4055 McPherson Lab, 
140 West 18th Avenue, Columbus, OH 43210, USA}
\altaffiltext{9}{Johns Hopkins University, 3400 N. Charles St., Baltimore, MD 21218, USA}
\altaffiltext{10}{Department of Physics, Drexel University, 3141 Chestnut Street, 
Philadelphia, PA 19104, USA}
\altaffiltext{11}{Department of Astronomy and Astrophysics, University of Chicago, 
Enrico Fermi Institute, 5640 South Ellis Avenue, Chicago, IL 60637, USA}
\altaffiltext{12}{Department of Astronomy and Astrophysics, The Pennsylvania State 
University, University Park, PA 16802, USA}

\begin{abstract}
We report the discovery of the two-image gravitationally lensed 
quasar SDSS J133222.62+034739.9 (SDSS J1332+0347) 
with an image separation of $\Delta \theta=1\farcs14$. 
This system consists of a source quasar at $z_s=1.445$ and a lens galaxy at $z_l=0.191$. 
The agreement of the luminosity, ellipticity and position angle of the lens galaxy 
with those expected from lens model confirms the lensing hypothesis. 

\end{abstract}


\section{Introduction}\label{sec:intro}

Strong gravitational lensing is a powerful probe of the large scale 
properties of the universe \citep{kochanek06}. 
The well-known physics makes it straightforward to use lensing 
in cosmological estimates of the dark energy \citep{turner90,fukugita90} 
and the Hubble constant \citep{refsdal64}.
The gravitational lensing effect depends solely on the mass density of the field, 
it is also an ideal tool for studies of dark matter in 
galaxies and clusters of galaxies \citep[e.g.,][]{oguri04a}. 
The Cosmic Lens All Sky Survey \citep[CLASS;][]{myers03,browne03}, which was conducted 
in the radio band, is the largest current statistical sample of lensed compact sources. 
However, the CLASS has only 13 lensed radio sources in its well-defined 
statistical sample. 
Larger lens samples are critical to using lensed quasars for cosmological and 
astrophysical tests. 

Optical surveys for lensed quasars are both complementary to radio surveys like 
the CLASS and likely to produce larger lens samples since only one-tenth of quasars 
are radio-loud \citep{ivezic02}. 
Our Sloan Digital Sky Survey Quasar Lens Search \citep[SQLS;][]{oguri06} is 
based on the Sloan Digital Sky Survey \citep[SDSS;][]{york00}, and its catalog of 
roughly 100,000 spectroscopically-confirmed quasars \citep[e.g.,][]{schneider05}. 
So far, the SQLS has succeeded in discovering 14 new lensed quasars 
\citep*{inada03a,inada03b,inada03c,inada05,inada06,johnston03,morgan03,pindor04,
pindor06,oguri04b,oguri05,burles06}, as well as recovering many previously known 
lensed quasars \citep{walsh79,weymann80,surdej87,bade97,oscoz97,schechter98,morgan01}. 
Indeed, the number of SDSS-discovered lensed quasars is already a significant 
fraction of all known quasar lenses ($\sim 80$\ 
\footnote{see CASTLES website, http://cfa-www.harvard.edu/castles/}). 
The large number of lenses contained in the SQLS and its well-defined selection 
function means that the lens catalog provided by the SQLS will be very useful for 
statistical studies of strong lensing \citep{oguri06}. 

In this paper, we report the discovery of a gravitationally lensed
quasar, SDSS J133222.62+034739.9 (hereafter SDSS J1332+0347). 
The quasar was confirmed to be doubly imaged by a low-redshift early-type galaxy,  
from observations using the Subaru 8.2-meter telescope and the University of 
Hawaii 2.2-meter (UH88) telescope. This paper is organized as follows. 
The SDSS data for SDSS J1332+0347, including its selection, is described in 
\S \ref{sec:sdss}.
The follow-up images and spectra are shown in \S \ref{sec:image} and 
\S \ref{sec:spec}, respectively. In \S \ref{sec:model} we model the lens. 
Finally we summarize our results in \S \ref{sec:conc}. 
Throughout the paper we assume a cosmological model with the matter
density $\Omega_M=0.27$, cosmological constant $\Omega_\Lambda=0.73$,
and Hubble constant $h=H_0/100\ {\rm km\,sec^{-1}Mpc^{-1}}=0.7$. 
Magnitudes are measured in AB system. 

\section{SDSS Data}\label{sec:sdss}

The SDSS is a survey to conduct both optical imaging and
spectroscopy. It covers about 10,000 square degrees of the sky approximately
centered on the North Galactic Cap with the dedicated wide-field
2.5-m telescope \citep{gunn06} at the Apache Point Observatory in New
Mexico, USA. The five broad-band photometric data are taken in imaging surveys 
\citep{fukugita96,gunn98,lupton99,tucker06} and are reduced automatically 
by the photometric pipeline \citep{lupton01}. From the magnitudes, colors, and 
morphologies of each object, quasar and galaxy candidates are selected 
as spectroscopic targets \citep{eisenstein01,richards02,strauss02}, 
and then assigned for spectroscopic observations by the tiling algorithm 
\citep{blanton03}. The spectra cover the wavelength range from 
3800{\,\AA} to 9200{\,\AA} with a resolution of $R\sim1800$. 
The astrometric accuracy is better than $\sim 0\farcs1$ rms
\citep{pier03} and the photometric zeropoint errors are less than $\sim 0.03$ 
magnitude over the entire survey area \citep{hogg01,smith02,ivezic04}. 
Most of the data have already become publicly available
\citep{stoughton02,abazajian03,abazajian04,abazajian05,adelman06}. 

SDSS J1332+0347 was selected as a lensed quasar candidate from the SDSS quasar sample, 
using the same selection algorithm as that described in \citet{inada03a}. 
In this algorithm, the morphological parameters of each quasar are used to select 
lensed quasar candidates. 
SDSS J1332+0347 was also selected by an improved algorithm \citep{oguri06} 
whose selection function has been extensively tested, and it is in the well-defined 
source quasar sample of the SQLS for lensed quasar statistics. 

\begin{figure}
\epsscale{.7}
\plotone{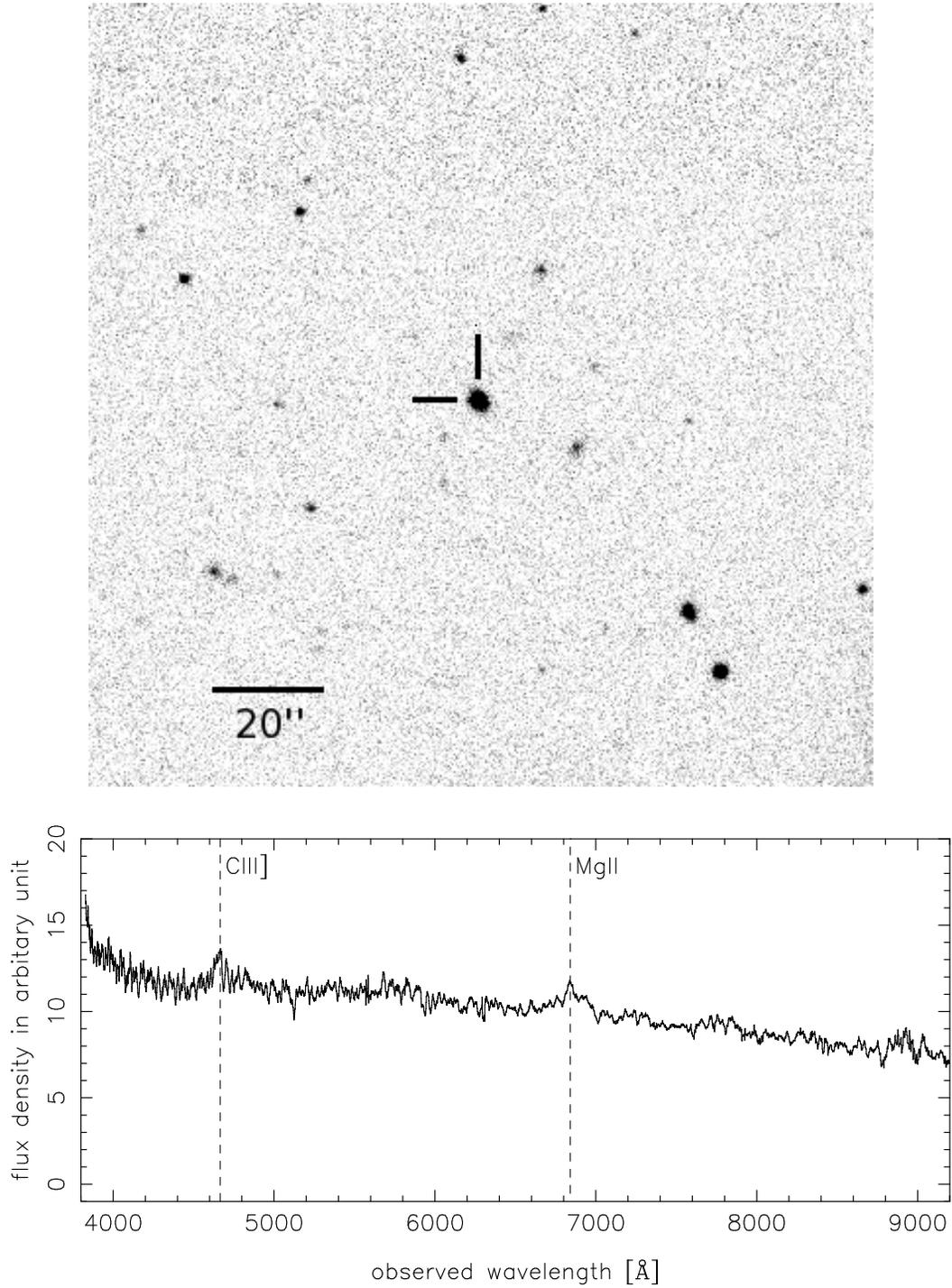}
\includegraphics[scale=0.6,angle=270]{f1b.eps}
\caption{upper panel: The wide-field SDSS $i$-band image of SDSS J1332+0347. The pixel 
scale is $0\farcs396\ {\rm pixel^{-1}}$. In the image, North is up and East is left. 
lower panel: The SDSS spectrum of SDSS J1332+0347. 
\label{fig:sdss_i}}
\end{figure}

Figure \ref{fig:sdss_i} shows the SDSS $i$-band image of the field 
under the seeing of $1\farcs6$. 
The SDSS Point Spread Function (PSF) magnitudes and the quasar redshift 
are summarized in Table \ref{tab:sdss}. 
The SDSS images indicate that the object is more 
extended than a single star, making it a good lens candidate 
and a target for higher angular resolution observations. 

\section{Imaging Observations}\label{sec:image}

We obtained a 90 sec $i$-band image of SDSS J1332+0347 with the Suprime-Cam 
\citep{miyazaki02} on the Subaru 8.2-meter telescope at Mauna Kea 
in seeing $0\farcs5$ on 2003 May 28. 
The $i$-band image was reduced in a standard way for the Suprime-Cam 
data using the NEKO software \citep{yagi02} and the SDFRED package \citep{ouchi04}. 
The absolute flux was calibrated using nearby stars in the SDSS catalogs. 
The small differences between the SDSS and Suprime-Cam $i$-band response 
functions are not important here. 
We also acquired a 720 sec $H$-band image of this object with the QUick near-InfraRed 
Camera (QUIRC) on the UH88 telescope 
in $0\farcs8$ seeing on 2005 February 20. 
The $H$-band image was reduced in a standard way using IRAF
\footnote{IRAF is the Image Reduction and Analysis Facility, a general purpose 
software system for the reduction and analysis of astronomical data. 
IRAF is written and supported by the IRAF programming group at the National Optical
Astronomy Observatories (NOAO) in Tucson, Arizona. NOAO is operated by
the Association of Universities for Research in Astronomy (AURA),
Inc. under cooperative agreement with the National Science Foundation.}, 
and calibrated using the standard star FS 23 \citep{hawarden01}. 

\begin{figure}
\epsscale{.8}
\includegraphics[scale=0.7]{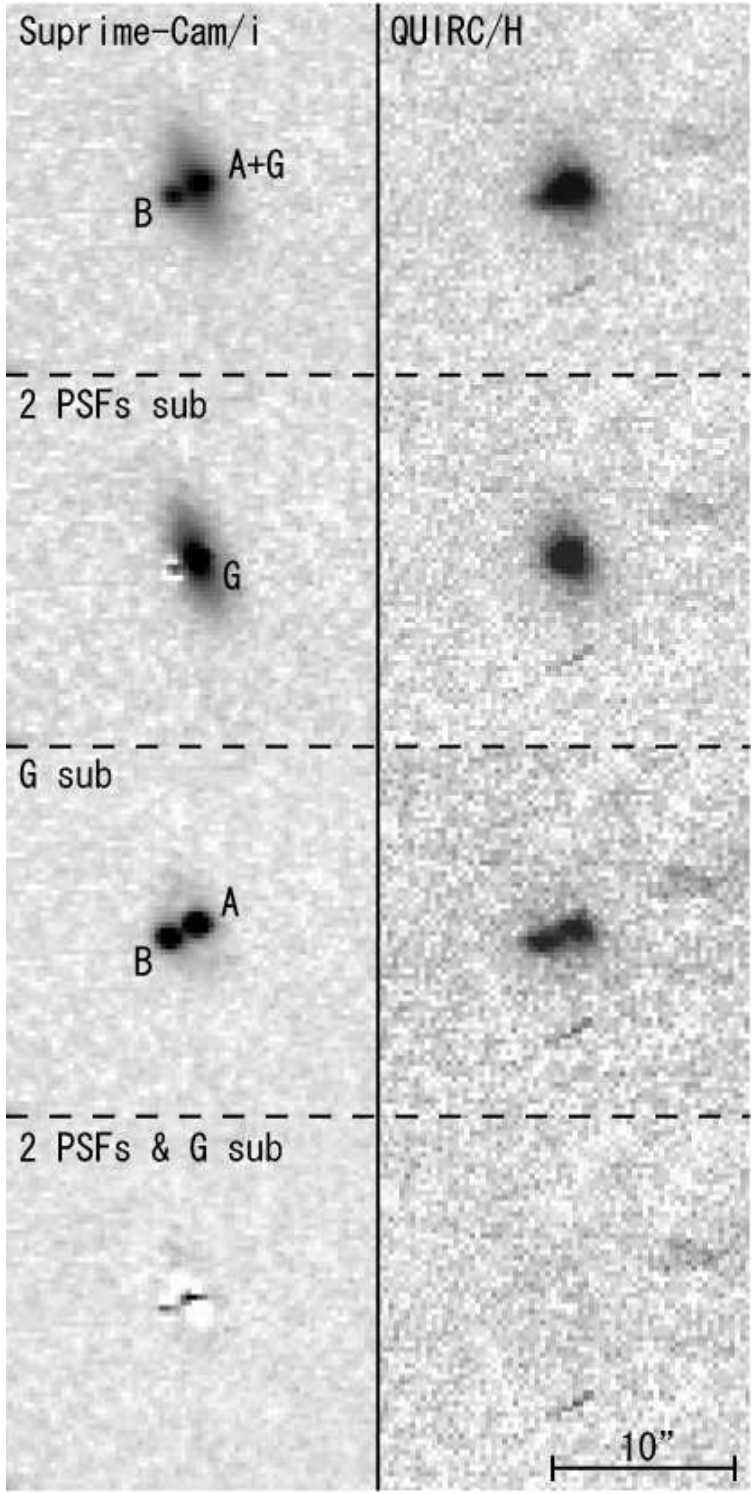}
\caption{\small{Follow-up images of SDSS J1332+0347. The optical $i$-band 
images taken with the Suprime-Cam on the Subaru telescope are shown in the left 
panels, while the near-infrared $H$-band images taken with the QUIRC on the 
UH88 telescope are displayed in the right panels. 
The original images, images after subtracting two PSF components, images after 
subtracting galaxy model, and images after subtracting both PSF and galaxy components 
are shown in the top, second, third, and bottom panels, respectively. 
In all panels, the box size is $20''$, and North is up and East is left. 
The pixel scales are $0\farcs202\ {\rm pixel^{-1}}$ 
(Suprime-Cam) and $0\farcs189\ {\rm pixel^{-1}}$ (QUIRC), respectively. 
\label{fig:images}}}
\end{figure}

The $i$ and $H$ images are shown in Figure \ref{fig:images}. 
Taking a closer look at the images, especially the $i$-band Suprime-Cam image, 
we see two compact sources bracketing an extended elliptical source. 
Indeed, models of the images using a public software, GALFIT \citep{peng02}, 
consisting of two point sources and an extended object with a de Vaucouleurs profile 
between them fit the data well. We used stars in the images as templates for the PSF 
in this procedure. We also show the observed galaxy after subtracting 
models for the two point sources, the observed stellar components after subtracting 
models for the galaxy, and the residuals after subtracting models for the point sources 
and the galaxy, in the second, third, and bottom panels of Figure \ref{fig:images}, 
respectively. 
We name the two stellar components A and B, where A is the brighter component that 
lies closer to the galaxy (separated by $\sim$0\farcs26), component G.  
The separation of the two stellar components derived from the $i$-band image is 
$\Delta\theta=1\farcs14$, and our relative astrometry is summarized in Table \ref{tab:1332}. 
The stellar components have similar colors of $i-H\sim1.6$ and $\sim1.8$~mag 
for A and B, respectively, 
while the galaxy G has a redder color of $i-H\sim 2.4$~mag. 
The photometry of the components is also summarized in Table \ref{tab:1332}. 
These astrometric and photometric properties are typical of known 
lensed quasar systems \citep[e.g.,][]{inada06}, 
and imply that both components A and B are lensed images of a quasar and 
component G is the lens galaxy. 

\section{Spectroscopic Observations}\label{sec:spec}

We carried out a spectroscopic observation of SDSS J1332+0347 using 
the Faint Object Camera And Spectrograph \citep[FOCAS;][]{kashikawa02} installed on 
the Subaru 8.2-meter telescope on 2003 June 20 with an exposure time of 900 sec 
in $0\farcs8$ seeing. 
The observation was conducted in 2$\times$2 on-chip binning mode, 
using a $0\farcs6$-width slit aligned along components A and B, 
with the grism 300B and the filter L600. 
This configuration provides a spectrum covering 3,900 \AA\ to 6,000 \AA\ 
with a spectral resolution of R$\sim700$, a spectral dispersion 
of 2.7 \AA\ pixel$^{-1}$ and a spatial pixel scale of $0\farcs208$. 
We used IRAF to analyze the spectra, extracting two traces corresponding to a 
blend of image A with the galaxy G and image B, respectively. 

\begin{figure}
\epsscale{.8}
\includegraphics[scale=0.65,angle=270]{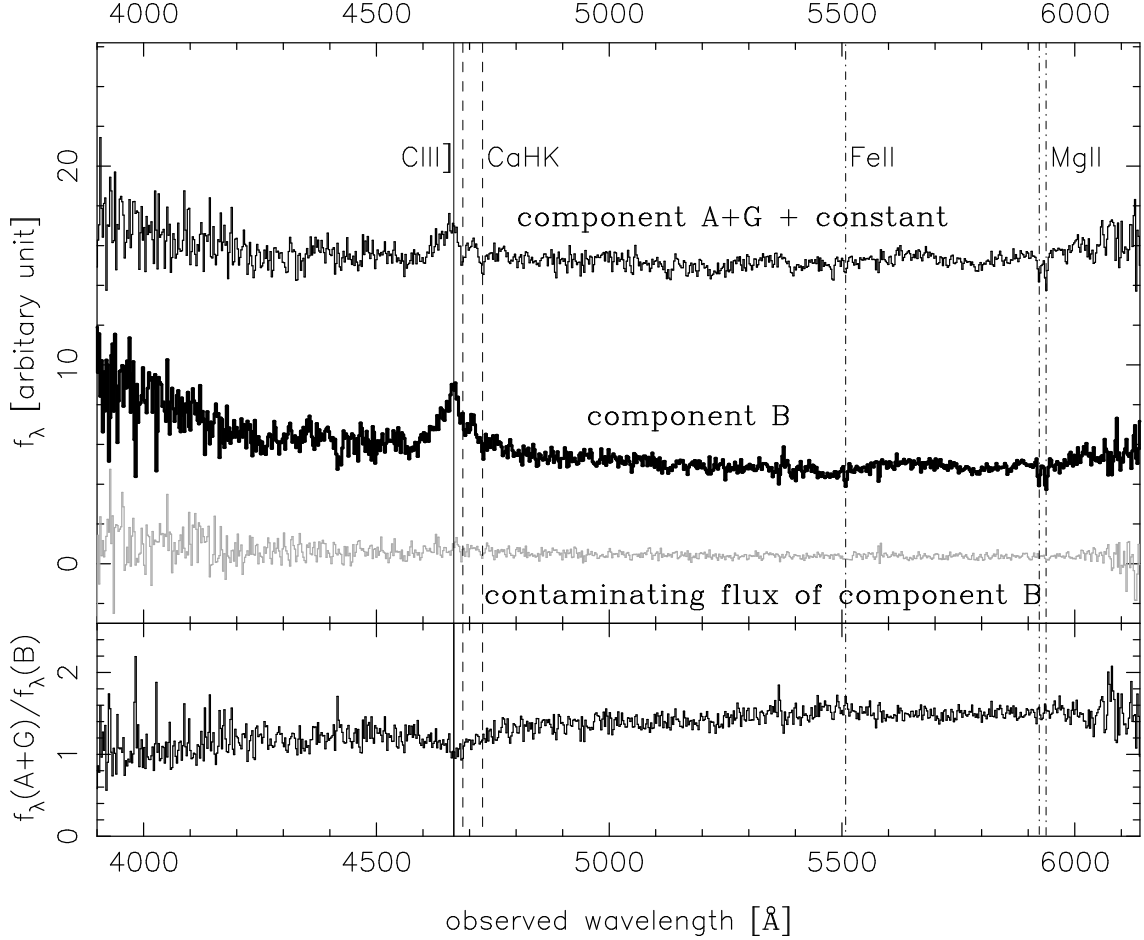}
\caption{Spectra (R$\sim$700) of components A+G (thin black solid line) and  B 
(thick black solid line) obtained with the Subaru/FOCAS in the upper panel. 
The spectrum of component A+G is shifted by constant to make it visible. 
A spectrum extracted at the same distance from component B as component A+G, 
but in the opposite direction, is also shown by the gray solid line. 
\ion{C}{3]} emission lines (vertical solid line) in components A and B at $z=1.445$, 
and \ion{Ca}{2} H and K absorption lines (vertical dashed line) of galaxy G 
at $z=0.191$ are detected. 
\ion{Fe}{2} and \ion{Mg}{2} absorption lines from an intervening 
system at $z=1.119$ are also detected (vertical dot-dashed line). 
The \ion{Mg}{2} absorption line doublet ratio of $1.0$ is typical of 
such systems \citep[e.g.,][]{steidel92}. 
The spectral flux ratio between components A+G and B is shown in the bottom panel. 
\label{fig:spec1}}
\end{figure}

The two spectra are shown in Figure \ref{fig:spec1}. 
We clearly see broad \ion{C}{3]} emission lines redshifted to $z=1.4$. 
On the red wings of the \ion{C}{3]} emission lines, we also see an absorption 
doublet at 4686 \AA\ and 4728 \AA, which we interpret as \ion{Ca}{2} H and 
K absorption lines redshifted to $z=0.191$. 
These strong absorption lines significantly distort the emission lines. 
It would be helpful to have a cleaner emission line in order to determine 
the quasar redshifts more accurately, but, the \ion{C}{3]} emission lines are 
the only strong emission lines in the FOCAS spectra. 
We estimated the quasar redshifts using three approaches; 
1) using the whole line profiles ($\lambda_{\rm obs}$), 
2) using only the line peaks ($\lambda_{\rm peak}$), and 
3) using the whole line profiles after masking and then interpolating through 
the absorption lines ($\lambda_{\rm mask}$).
The results, summarized in Table \ref{tab:caiii}, confirm that the two quasars 
have the same redshifts. 
We adopt procedure 3) for our standard result of $z=1.445$ as it is likely 
to be the most reliable. 
We also verified that the spectrum A+G has little contamination from quasar B 
by extracting a spectrum at the same distance from component B as component A+G 
but in the opposite direction. 
The spectrum, shown in Figure \ref{fig:spec1}, has no significant emission, 
and we conclude that the \ion{C}{3]} emission line in the spectrum A+G is 
due to component A. 
Furthermore, in the spectra of both components A+G and B, we can see \ion{Mg}{2} 
$\lambda\lambda$2796,2803 doublet absorption lines 
and a \ion{Fe}{2} $\lambda$2600 absorption line 
by an intervening object at $z=1.119$. 

\begin{figure}
\epsscale{.8}
\includegraphics[scale=0.65,angle=270]{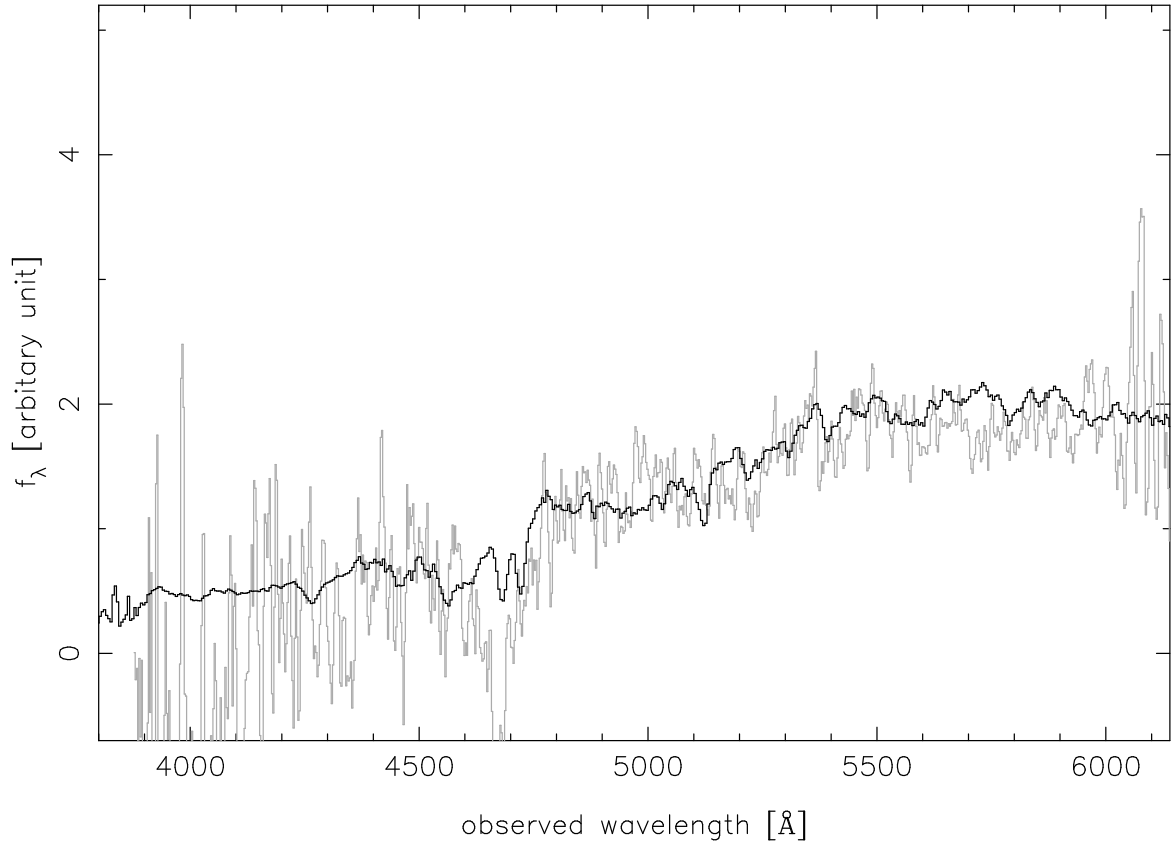}
\caption{A spectrum of component G (gray solid line) obtained 
by subtracting a scaled (multiplied by 1.127) spectrum of component B
from that of component A+G. 
The derived galaxy spectrum does not clearly show the \ion{Ca}{2} H and K 
absorption lines, but is consistent with a template spectrum of an elliptical galaxy 
\citep{kinney96} at $z=0.191$ (black solid line). 
This result indicates that components A and B have identical SEDs and redshifts. 
\label{fig:spec2}}
\end{figure}

The spectral flux ratio between components A+G and B, displayed in the bottom 
panel of Figure \ref{fig:spec1}, indicates that combination of A+G is 
redder than component B. 
If we scale the spectrum B by the $H$-band flux ratio of $1.127$ between 
components A and B, and subtract it from the spectrum A+G, then we obtain 
the residual spectrum shown in Figure \ref{fig:spec2}. 
While the spectrum is noisy, especially around the \ion{C}{3]} emission line 
and towards the blue wavelength region, the spectral energy distribution (SED) 
is consistent with a template spectrum of an elliptical galaxy \citep{kinney96} 
at the redshift of the \ion{Ca}{2} absorption feature, $z=0.191$. 
Given the strength of the absorption, we conclude that the SEDs of components A 
and B are identical and that component G is an elliptical galaxy at $z=0.191$. 

\section{Lens Model}\label{sec:model}

We used the {\it lensmodel} package \citep{keeton01} to fit a mass model 
to the observations of SDSS J1332+0347, assuming that the object is indeed 
a lensed quasar system. 
We assume the standard Singular Isothermal Ellipsoid (SIE) model 
characterized by eight parameters: the Einstein radius $R_{\rm E}$, 
ellipticity $e$, position angle $\theta_e$, position of 
the lensing galaxy, the position and flux of the source quasar. 
Since the number of observable constraints is also eight (the positions of A, B, 
and G, and the fluxes of A and B), the number of degrees of freedom is zero.
As expected, we were able to find a model that perfectly reproduces the
observables, $\chi^2\sim 0$. 
The results of the model are presented in Table~\ref{table:model}.  
We note that the best-fit values for the ellipticity and the position
angle, $e=0.53\pm0.07$ and $\theta_e=25^\circ\pm2^\circ$, are in good agreement 
with those observed for the galaxy G (derived using GALFIT in $i$-band image; 
see Figure \ref{fig:images}), of $e=0.69$ and 
$\theta_e=21^\circ$, respectively. 

Einstein radii $R_{\rm E}$ are related to the velocity dispersions
of lensing galaxies, so we can use the Faber and Jackson relation \citep{faber76} 
to estimate the luminosity of the lens galaxy. 
From the best-fit value $R_{\rm E}=0\farcs465$, we estimate that the apparent magnitude 
of the lens is $i=18.8$, assuming the correlation of velocity dispersions and 
magnitudes of early-type galaxies derived  by \citet{bernardi03}. This is again
in good agreement with the observed magnitude of the galaxy G, $i=18.6$
(Table \ref{tab:massmodel}), implying that galaxy G is responsible for
the most of the lens potential.

\section{Conclusions}\label{sec:conc}
We have presented the extensive follow-up observations of SDSS J1332+0347
using the Subaru telescope and UH88 telescope. 
We found a bright galaxy (component G) between
two stellar components (components A and B) with similar color in both the 
Subaru/Suprime-Cam $i$-band and UH88/QUIRC $H$-band images. 
Although the \ion{C}{3]} emission lines of the two stellar 
components are distorted by the \ion{Ca}{2} H and K absorption lines from the bright 
galaxy at $z_l=0.191$, we can still confirm that the two quasars are at the same 
redshift, $z_s=1.445$. 
Subtraction of a scaled spectrum of component B from the component A+G leaves 
a residual that is consistent with the spectrum of an elliptical galaxy redshifted 
to $z=0.191$. 
The observed luminosity, ellipticity and position angle of the bright galaxy 
are in good agreement with those expected from a standard lens model. 
All these results lead us to the conclusion that SDSS J1332+0347 is 
indeed a lensed quasar system, where a source quasar at $z_s=1.445$ is 
lensed by a bright lens galaxy at $z_l=0.191$ resulting in an image separation of 
${\Delta}{\theta}=1\farcs14$. 
One peculiar characteristic of the lens is that the brighter component A 
is closer to the lens galaxy than the fainter component B. 
We have no difficulty reproducing the flux ratio, and note that such {\it inverted} 
flux ratios were also observed and easily modeled in 
the two-image lens HE1104--1805 \citep{wisotzki93}. 
This discovery adds another object to the complete sample of
the SDSS lensed quasars that can be used to estimate the cosmological model.
The brightness of the lens galaxy also makes it a good candidate for dynamical 
observations; higher-resolution imaging would allow us to determine more accurately 
the brightness of the components in this lens system, and spectroscopic observations 
with better signal-to-noise ratios would also allow us to study the interstellar medium 
in the lens galaxy in detail. 


\acknowledgments

N.~I. is supported by JSPS through JSPS Research Fellowship
for Young Scientists. This paper is based in part on data collected at Subaru
Telescope, which is operated by the National Astronomical Observatory of
Japan. Use of the UH 2.2-m telescope for the observations is supported
by NAOJ. I.~K. acknowledges the support from Ministry of Education, Culture, 
Sports, Science, and Technology, Grant-in-Aid for Encouragement 
of Young Scientists (No. 17740139).
This work was supported in part by the Department of Energy contract 
DE-AC02-76SF00515.

Funding for the SDSS and SDSS-II has been provided by the Alfred P. Sloan Foundation, 
the Participating Institutions, the National Science Foundation, the U.S. Department 
of Energy, the National Aeronautics and Space Administration, the Japanese Monbukagakusho, 
the Max Planck Society, and the Higher Education Funding Council for England. 
The SDSS Web Site is http://www.sdss.org/.

The SDSS is managed by the Astrophysical Research Consortium for the Participating 
Institutions. The Participating Institutions are the American Museum of Natural History, 
Astrophysical Institute Potsdam, University of Basel, Cambridge University, Case Western 
Reserve University, University of Chicago, Drexel University, Fermilab, the Institute 
for Advanced Study, the Japan Participation Group, Johns Hopkins University, 
the Joint Institute for Nuclear Astrophysics, the Kavli Institute for Particle 
Astrophysics and Cosmology, the Korean Scientist Group, the Chinese Academy of Sciences 
(LAMOST), Los Alamos National Laboratory, the Max-Planck-Institute for Astronomy (MPIA), 
the Max-Planck-Institute for Astrophysics (MPA), New Mexico State University, 
Ohio State University, University of Pittsburgh, University of Portsmouth, Princeton 
University, the United States Naval Observatory, and the University of Washington.

\clearpage

\begin{deluxetable}{cccccc}
\tabletypesize{\normalsize}
\tablecaption{SDSS J1332+0347: SDSS PHOTOMETRY AND REDSHIFT\label{tab:sdss}}
\tablewidth{0pt}
\tablehead{
\colhead{$u$\tablenotemark{a}} &
\colhead{$g$\tablenotemark{a}} & \colhead{$r$\tablenotemark{a}} & 
\colhead{$i$\tablenotemark{a}} & \colhead{$z$\tablenotemark{a}} & 
\colhead{Redshift\tablenotemark{b}} }
\startdata
19.12{$\pm$}0.03 & 18.61{$\pm$}0.03 & 18.26{$\pm$}0.02 & 17.95{$\pm$}0.02 
& 17.84{$\pm$}0.02 & 1.438{$\pm$}0.003 \\
\enddata
\tablenotetext{a}{PSF magnitudes from the SDSS imaging data.} 
\tablenotetext{b}{quasar redshift from the SDSS spectrum.} 
\end{deluxetable}

\begin{deluxetable}{crrrr}
\tablewidth{0pt}
\tablecaption{SDSS J1332+0347: ASTROMETRY AND PHOTOMETRY\label{tab:1332}}
\tablehead{\colhead{Object} & \colhead{$x$[arcsec]\tablenotemark{a}} &
 \colhead{$y$[arcsec]\tablenotemark{a}} &  
 \colhead{$i$\tablenotemark{b}} & \colhead{$H$\tablenotemark{b}}} 
\startdata
A & $0.000\pm0.006$  & $0.000\pm0.006$  & $19.28\pm0.03$ & $17.72\pm0.17$ \\
B & $-1.024\pm0.006$ & $-0.501\pm0.006$ & $19.66\pm0.02$ & $17.85\pm0.05$ \\
G & $-0.228\pm0.013$ & $-0.119\pm0.013$ & $18.64\pm0.02$ & $16.28\pm0.04$ \\
\enddata
\tablecomments{All the values are the outputs of the GALFIT procedure for 
SDSS J1332+0347. }
\tablenotetext{a}{The positive directions of $x$ and $y$ are West and
 North, respectively. The values are measured in the Subaru Suprime-cam
 $i$-band image.} 
\tablenotetext{b}{The errors do not include the absolute calibration
 uncertainty.} 
\label{tab:massmodel}
\end{deluxetable}
\begin{deluxetable}{ccccccc}
\tablewidth{0pt}
\tablecaption{SDSS J1332+0347: CENTRAL WAVELENGTHS OF \ion{C}{3]}
\label{table:specline}}
\tablehead{
\colhead{Object} 
& \colhead{$\lambda_{\rm obs}$}
& \colhead{$z_{\rm obs}$}
& \colhead{$\lambda_{\rm peak}$}
& \colhead{$z_{\rm peak}$}
& \colhead{$\lambda_{\rm mask}$}
& \colhead{$z_{\rm mask}$}
}
\startdata
A+G & $4653.49$ & $1.4380$ & $4661.11$ & $1.4420$& $4666.23$ & $1.4447$\\
B   & $4663.26$ & $1.4431$ & $4666.08$ & $1.4446$& $4666.25$ & $1.4447$\\
\enddata
\label{tab:caiii}
\end{deluxetable}

\begin{deluxetable}{cc}
\tablewidth{0pt}
\tablecaption{SDSS J1332+0347: LENS MODEL\label{table:model}}
\tablehead{\colhead{Parameter} & \colhead{Value} }
\startdata
Einstein radius                  & $R_{\rm E}=0\farcs465$ \\
Ellipticity                      & $e=0.53\pm0.07$ \\
Position angle\tablenotemark{a}  & $\theta_e=25^\circ\pm2^\circ$ \\
Total magnification              & $\mu_{\rm tot}=4.1$ \\ 
Time delay                       & $\Delta t=6.7h^{-1}$day \\
\enddata
\tablenotetext{a}{Each position angle is measured East of North.}
\end{deluxetable}

\clearpage

\end{document}